\documentclass[pre,aps,a4paper,oneside,fleqn,twocolumn,showpacs,showkeys]{revtex4}
\usepackage[latin1]{inputenc}
\usepackage[]{natbib}
\usepackage{graphicx}
\usepackage{dcolumn}
\usepackage{amsmath}
\begin{document}
\title{The geometrical pattern of the evolution of cooperation in the Spatial Prisoner's Dilemma: an intra-group model}

\author{
\firstname{Ricardo} Oliveira dos Santos  \surname{Soares} 
%\email{ricardosoares@aluno.ffclrp.usp.br} 
and 
\firstname{Alexandre} Souto \surname{Martinez}
\email{asmartinez@ffclrp.usp.br}
%\homepage{http://fisicamedica.com.br/martinez/}
}
\affiliation{Faculdade de Filosofia, Ci\^encias e Letras de Ribeir\~ao Preto,
Universidade de S\~ao Paulo \\ Av. Bandeirantes, 3900 \\ 14040-901, 
Ribeir\~ao Preto, SP, Brazil.}
\date{\today}

\begin{abstract}
The Prisoner's Dilemma (PD) deals with the cooperation/defection conflict between two agents. 
The agents are represented by a cell of $L \times L$ square lattice. 
The agents are initially randomly distributed according to a certain proportion $\rho_c(0)$ of cooperators.
Each agent does not have memory of previous behaviors and plays the PD with eight nearest neighbors and then copies the behavior of who had the greatest payoff for next generation.
This system shows that, when the conflict is established, cooperation among agents may emerge even for reasonably high defection temptation values.
Contrary to previous studies, which treat mean inter-group interaction, here a model where the agents are not allowed to self-interact, representing intra-group interaction, is proposed.
This leads to short time and asymptotic behaviors similar to the one found when self-interaction is considered.
Nevertheless, the intermediate behavior is different, with no possible data collapse since oscillations are present.
Also, the fluctuations are much smaller in the intra-group model. 
The geometrical configurations of cooperative clusters are distinct and explain the $\rho_c(t)$ differences between inter and intra-group models.
The boundary conditions do not affect the results. 
\end{abstract}

\maketitle

%\tableofcontents

%4 \section{Introduction}

Although it has been always observed in natural systems, spontaneous cooperation has not a foothold within the Darwinian Evolutionary theory, which frequently focuses to direct competition. 
No evolutionary modification over a species can be selected by the natural selection if the new trait is exclusively beneficial to other species. 
Nevertheless, there can be indeed a selective production of directly harmful structures to other animals, such as the viper's hooks~\cite{darwin}. 
It may also be suggested that Dawkins' emphasis on selfish genes~\cite{dawkins} comes to give a new perspective on the role of competition since it envisages even the possibility of competition inside the genome.
On the other hand, cooperation is favored in group selection where although the detriment to the fitness of individuals who, for instance, express a given gene, it may be advantageous to the group (demes) which the individuals are members~\cite{boorman}. 

The conflict between cooperation and competition in the game theory, especially in the context of the  Prisoner's Dilemma (PD) problem, also known as ``the tragedy of the commons''~\cite{hardin_1968} when defection dominates, has been greatly investigated. 
Consider the following rule~\cite{nowak1992}: two players can have payoff of either $R = 1$ (reward) or $P = 0$ (punishment), if both cooperate or both defect, respectively. 
If one of them defect, the defector has a payoff of $T$ (temptation) and the cooperator of $S = 0$ (sucker).
The conflict is set up with cooperation being the best global strategy and defection the best individual strategy, which occurs for~\cite{axelrod1984}: $T > R > P > S$ and $2R > S + T$.
For the values employed here, without any harm, the condition $P > S$ has been relaxed and the conflict range is: $1<T<2$.
Temptation is the only free parameter in this model. 
If one assigns a state $\theta = 0$ ($\theta = 1$) for defection (cooperation)~\cite{schweitzer_2002}, the payoff of agent $i$, in state $\theta_i$, given that agent $j$ is in the state $\theta_j$, is: $J(\theta_i|\theta_j) = \theta_i \theta_j + T (1 - \theta_i \theta_j)\theta_j$. 

A major advance in the comprehension of cooperation-competition conflict has been made with the consideration of one memory step for the agent when the same agents play PD sucessively (Iterated Prisoner's Dilemma). 
The cooperation has first emerged in this system as a component of an optimum strategy in either with tit-for-tat in deterministic environment~\cite{axelrod1981,axelrod1984} [$\theta_i(t+1) = \theta_j(t)$, where $t$ is the time step] or win-stay lose-shift, also known as the Pavlov strategy, in stochastic environment~\cite{nowak1993} \{$\theta_i(t+1) = \theta_i(t)$ if $J[\theta_i(t)|\theta_j(t)] \ge J[\theta_i(t-1)|\theta_j(t-1)]$ or  $\theta_i(t+1) = 1 - \theta_i(t)$, otherwise\}.

Nowak and May~\cite{nowak1992,nowak1994} considered the interaction among several agents, who play PD repetitively, i.e., an $N$-agent game. 
This analysis of the PD, including the spatial component (SPD), has validated the argument that cooperation could emerge as a stable strategy when the dilemma is played among several agents dynamically.
This model is interesting when dealing with a whole conglomerate of individuals (group), each represented by a cell of a square lattice~\cite{nowak1992}.
This conglomerate can be either in a cooperative or defective state and can play the PD with itself, setting up an environment of interacting groups - an inter-group model.
We stress that in this case self-interaction is well justified for sucessive time steps in demes and that cooperation states may emerge from group selection models as Refs.~\cite{fontanari_epjb_1999,fontanari_pre_2000} sugest.  
Also, the SPD has been played on more realistic structures, which mimic social relations, such as the small world networks~\cite{kim}, disordered lattices~\cite{vainstein_2001} and random graphs~\cite{duran_2003}, increasing the scope of application of single PD  which ranges from gene polymorphism in yeast~\cite{greig}, intra-host competition in RNA virus~\cite{turner1,turner2} to predator inspection in fishes~\cite{dugatkin_1989,dugatkin_1991,milinski}.

In the SPD there exist three main regimes as a function of $T$, where temptation is not the only evolutionary stable strategy~\cite{smith_1982}. 
For $T \sim 1$ ($T \sim 2$), the proportion of cooperators $\rho_c$ is stationary and majority (minority).
For intermediate values of $T$, $\rho_c$ is non-stationary even presenting chaos~\cite{nowak1992}.
However, each main regime may be divided into regions where $\rho_c(T)$ change values according to the number of interacting neighbors $\alpha$ and number of interacting cooperators $0 \le s \le \alpha$.
If agent $i$ has $s_i$ cooperative agents around, his/her payoff is~\cite{duran_2003}:  $g_{\theta_i}^{(s_i)} = [T - (T - 1)\theta_i] s_i$. 
This leads to: $g_1^{(s)} = s$ and $g_0^{(s)} = T s$, since $T > 1$, $g_0^{(s)} > g_1^{(s)}$ and     $g_{\theta}^{(s)} \ge g_{\theta}^{(s-1)}$.
The transitions are given by~\cite{duran_2003}: $T_{n,m} = (\alpha - n)/(\alpha - n - m)$, since $1 < T < 2$, then  $0 \le n < \alpha$ and $1 \le m \le \mbox{int}[(\alpha - n - 1)/2]$. 
These values are shown in Table~\ref{table_1}.
Also as pointed out by Schweitzer~\emph{et al.}~\cite{schweitzer_2002} and Dur\'an and Mulet~\cite{duran_2003}, it is interesting to view the system as two populations that invade each other instead of agents switching states as a function of time.

\begin{table}
\begin{center}
\begin{tabular}{cc|c|c|c|c|c|c|c}
\hline
     & $\alpha$    & $9 \rightarrow$    & $ 8 \rightarrow$    & $ 7 \rightarrow$    & $ 6 \rightarrow$    & $ 5 \rightarrow$    & $ 4 \rightarrow$    & $ 3$                \\  
\hline     
 $m$ & $n$ & 0                           & 1                           & 2                           & 3                           & 4                           & 5                           & 6                           \\
\hline
 \hline
  1  &     & $9/8$                       & $8/7$               & $7/6 $ & $6/5 $ & $5/4 $ & $4/3 $ & $3/2$  \\
\hline 
  2  &     & $9/7$                       & $8/6$               & $7/5 $ & $6/4 $ & $5/3 $ &  2     &                             \\
\hline
  3  &     & $9/6$                       & $8/5$               & $7/4 $ &   2    &                             &                             &                             \\
\hline  
  4  &     & $9/5$                       &  2                  &                 &                             &                             &                             &                             \\
\hline
\end{tabular}
\end{center}
\caption{The Prisoner's Dilemma transitions as a function of neighborhood  $\alpha$ are given by: $T_{n,m} = (\alpha - n)/(\alpha - n - m)$ with $1 < T < 2$, $0 \le n < \alpha$ and $1 \le m \le \mbox{int}[(\alpha - n - 1)/2]$.~\cite{duran_2003}}
\label{table_1}
\end{table}

Nevertheless, if one considers an intra-group model, where the lattice represents a group and the cells are thought as individuals, one can hardly explain this self-interaction as a viable social and biological behavior.
This increases the scope of application of the SPD in sociophysics~\cite{staufer_2004,galam_2004} and agent-based models of econophysics~\cite{bouchaud_2002,anteneodo:1:2002}. 

In this paper we compare the intra ($\alpha = 8$) and inter-group ($\alpha = 9$) models. 
We show that self-interaction produces a bias to cooperative states leadingto more stable cooperation clusters which increase  fluctuations of the cluster size modifying the  dynamics of the system.
We start describing the model then extend the boundary analysis to the cooperation cluster and finally we present and interpret the results obtained by numerical simulation.

%4 \section{Model Description}

Consider an initial randomly distributed proportion $\rho_c(0)$ of agents who cooperate in the $L \times L$ cells of a square lattice. 
The initial distribution of collaborators is the only stochasticity considered and the evolution of the system is  entirely deterministic afterwards. 
At each generation, each agent plays the PD with the first and second neighbors totalizing $\alpha = 8$.  
The given agent compares his/her own payoff with the considered neighbor ones and, for the next generation, copies the agent state $\theta$ of whom had the highest payoff. 
The scores of the agents are not cumulative; all the payoffs are reset to zero after each PD round (``one-shot game'').     
The case where agents self-interact and interact with only the nearest neighbors ($\alpha = 5$) is treated in Ref.~\cite{schweitzer_2002}. 

\begin{figure}
\begin{center}
\includegraphics[angle=-90,width=\columnwidth]{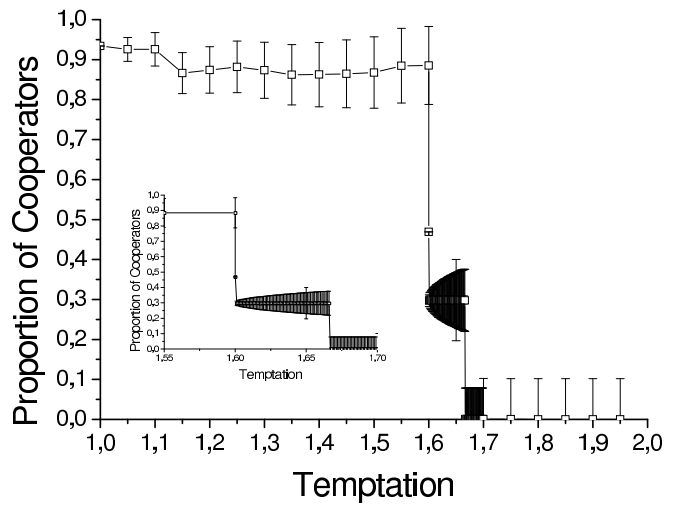}
\end{center}
\caption{Graphical representation of the asymptotic values (1000 generations) of an initial proportion of cooperation $\rho_c(0) = 0.60$ in 1000 clone groups of non-self-interacting agents in square lattice ($L= 200$) as a function of temptation values using periodic boundary condition. 
This pattern is consistent with the transition values with $\alpha = 8$ in Table~\ref{table_1}.
The error bars refer to the standard error of the mean.
{\bf Inset:} Amplification of the cooperation/defection coexistence region.}
\label{fig1}
\end{figure}

In the system where the agents self-interact ($\alpha = 9$), the cooperation/defection coexistence region is in the interval $9/5<T<2$.
For not self-interacting agents ($\alpha = 8$), this region is: $8/5 < T < 5/3$ (see Fig.~\ref{fig1} and Table~\ref{table_1}).  
Notice that contrary to inter-group model, the defection regime is present near the middle of the conflict region ($1 < T < 2$).
Neglecting self-interaction does not drastically change none of the two regimes (stationary, non-stationary)  nor the pattern found by Nowak \& May~\cite{nowak1992}, but only shifts them to lower temptation values. 

% \subsection{Boundary Effect}

To understand the dynamics and to quantify payoffs, let us first classify the agents according to the position of the cell in the $L \times L$ square lattice.
There are: $(L-2)^2$ of inner (bulk) agents who have 4 first neighbors and 4 second neighbors ($\alpha_b = 8$); $4(L-2)$ of surface agents who have 3 first neighbors and 2 second neighbors ($\alpha_s = 5$) and $4(L-2)$ of edge agents have 2 first neighbors and 1 second neighbor ($\alpha_e = 3$). 
As the order of adjacency is decreased, the number of first neighbors  decrease arithmetically while the number of second neighbors is halved (geometrically). 
Considering the variables $
I_{ext}  =  4 (L - 1)$, $I_{int}  =  2 (L - 1)(L - 2)$ and $I_{cross}  =  2 (L - 1)^2$ and adding them, one obtains the number of times the PD is played in each generation for fixed boundary condition (FBC) $N_{f} = I_{ext} + I_{int} + I_{cross} = 4 \; (L-1) \; ( L- 1/2)$, 
and for periodic boundary conditions (PBC)
$N_{p} = 2[ 2(L-1) ( L - 1/2 ) + L + 1 ] = N_{f} + 2L + 2$.
This quantifies  the effect of the PBC on the group. 
For $L \gg 1$, $N_p \simeq N_f \simeq 4L^2$, the remaining difference between PBC and FBC being of order $L$.

%\subsection{Cluster of Cooperative Cells}

The boundary effect of cooperative cluster agents provides some understanding about the dynamics  of the system.  
The idea of cooperation adjacency reveals several scenarios going from the bulk agents, who have payoff $8$ units (due to 8 cooperative agents around), surface agents with payoff 5 and convex edge agents with payoff 3 while the payoff of the surrounding defector agents are $T$. 
The convex edge agents are the more unstable ones.    
Edge agents can be either convex, if the agent have 2 nearest and 1 next-nearest cooperators or concave, if the agent have 4 nearest and 3 next-nearest cooperators.  

\begin{figure}
\begin{center}
\includegraphics[angle=-90,width=.9\columnwidth]{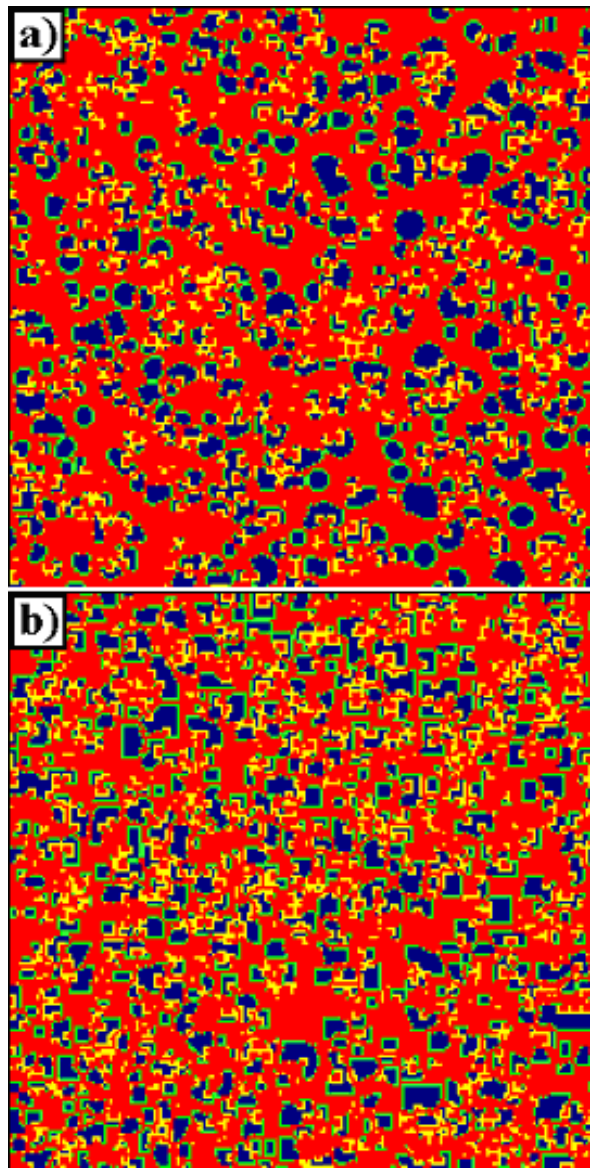}
\end{center}
\caption{One realization of cooperator and defectors with $T_{nsi} = 1.65$ and $T_{si} = 1.95$ when self-interaction is neglected {\bf (a)} and considered {\bf (b)}, as $t \rightarrow \infty$, with periodic boundary condition. 
The initial values are $\rho_c(0) = 0.60$ for a square lattice of size $L = 200$. 
The following color coding has been used: blue (dark gray), the agent is a cooperator; green (light gray), the agent is a former defector; red (gray), the agent is a defector; yellow (very light gray),  the agent is a former cooperator. 
Without self-interaction (top) cooperative clusters grow as a round-shaped forms while with self-interaction (bottom) these clusters grow as square-shaped forms.}
\label{fig2}
\end{figure}

The deterministic dynamics forbids inner (bulk) agents to switch state if all the nearest and next-nearest neighbors have the same state (bulk cell of a cooperative cluster, for instance). 
The cooperative cluster conformations start to be altered in the convex edge cells, because of the smallest payoff leading them to switch state and producing new edge cells which, again are more susceptible to switch state and permit the cooperation cluster to be invaded.  
As illustrated in Fig.~\ref{fig2}, cooperation clusters evolve differently when self-interaction is considered or not.
If self-interaction is considered, the cooperative clusters are essentially squares, which have 4 convex edges. 
On the other hand, when self-interaction is neglected, the cooperation clusters are essentially spherical, having more convex edges which diminishes the cluster stability. 
In Fig.~\ref{fig2}, in order to have around the same cooperation cluster density, it is necessary to have smaller temptation values $T$ when self-interaction is neglected. 

%4 \section{Simulation Results and Analysis}

\begin{figure}
\begin{center}
\includegraphics[angle=-90,width=\columnwidth]{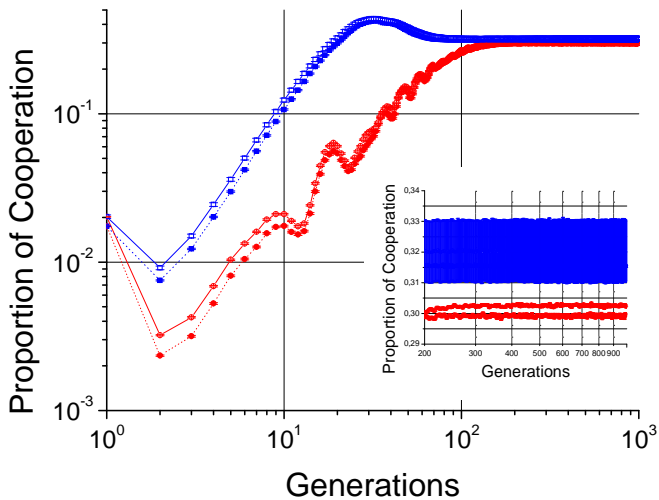}
\end{center}
\caption{Proportion of cooperators $\rho_c$ as a function of time $t$ for $T_{si} = 1.95$ and $T_{nsi} = 1.65$ for inter (below) and  intra-group (above) with fixed (line) and periodic (dotted line) boundary conditions. 
The initial values are $\rho_c(0) = 0.60$ for a $L = 200$ lattice and $M = 1000$ realizations. {\bf Inset:} Same but with $200 < t < 1000$. 
The fluctuations are much smaller in the inter-group model.}
\label{fig3}
\end{figure}

A comparison between the evolution of cooperators proportion $\rho_c(t)$, with and without  self-interaction as a function of time (generations) $t$, is shown in Fig.~\ref{fig3} for fixed and periodic boundaries conditions.
Starting with $\rho_c(0) = 0.60$, both systems react in the first generation by lowering $\rho_c(t)$ and then increasing it by the formation of cooperation clusters. 
The seed to form a cooperation cluster is when four nearest neighbors cells are in a cooperative state. 
This increase of $\rho_c(t)$ occurs up to around ten generations for both models. 
We point out that $\rho_c^{(si)}(t) \sim 5 \rho_c^{(nsi)}(t)$, for $1 \le t \le 10$.    
The intermediate regime is the most interesting one since the differences between the inter and intra-group models are striking, while $\rho_c^{(si)}(t)$ is smooth, $\rho_c^{(nsi)}(t)$ oscillates. 
Finally, after about 100 generations both systems stabilize and the proportion of cooperation is $\rho_c^{(si)}(\infty)=0.318$ and $\rho_c^{(nsi)}(\infty)=0.299$.
These values are very close but they are significantly distinct, once they persist indefinitely as stationary values (see inset of Fig.~\ref{fig3}).
The error bars, which are the standard error of the mean, are one order of magnitude smaller for the intra-group model.

In the intermediate regime, from 10 up to 100 generations, $\rho_c^{(si)}(t)$ grows rapidly presenting a great bump ($t \sim 25$) before stabilization. 
On the other hand, $\rho_c^{(nsi)}(t)$ grows less rapidly and presents smaller, but more frequent, bumps (oscillatory behavior). 
The reason for these different behaviors can be understood with the consideration of geometrical structures of the cooperation clusters, which implies in different cooperating cluster stabilities. 
As we have seen, these clusters are square-like and circle-like when self-interaction is considered or neglected, respectively, with the former one being more stable than the latter. 

When agents self-interact, the initial cooperative clusters grow, increasing $\rho_c^{(si)}(t)$ until they crash with each other (and/or touch the boundaries for FCB case) producing  numerous less stable cooperation clusters. 
This crash generates the great bump (Fig.~\ref{fig3}) and may produce larger fluctuation when compared to the intra-group model (see inset of Fig.~\ref{fig3}). 
When the agents do not self-interact, the clusters growing behavior is different, because they are rounded, having more convex edge cells, so that the clusters grow until they reach a critical size and then break in parts which are the seeds for new cooperation clusters.
These clusters have a notable ability to self-disperse in more homogeneous smaller growing islands. 
This mechanism produces the transient oscillations and smaller fluctuations observed (see inset of Fig.~\ref{fig3}) found for the inter-group model.

%4 \section{Conclusion}

We have considered (inter group) and neglected (intra group) the self-interaction of agents in the spatial DP game.
For short and long times,  both models present about the same average behavior.
Nevertheless, the fluctuations for intergroup are greater (one order of magnitude) for the intragroup case.
In the intermediate regime they drastically differ due to the geometrical shape of the cooperation cluster which are typically square-like for inter group and circle-like for intra-group models. 
The overall effect of self-interaction is to favor cooperation in all steps of evolution and increase fluctuation due to the more stability of the cooperation clusters.

%\acknowledgements

ROSS acknowledges the support of the Brazilian agencies CNPq and FAPESP (03/09884-7) and ASM acknowledges the support of CNPq (305527/2004-5) while this study has been under development.
We would like to thank O. Kinouchi (who called our atttention to Refs.~\cite{fontanari_epjb_1999,fontanari_pre_2000}), C. Rodrigues Neto, E. E. S. Ruiz and H. F. Terenzi for very stimulating and fruitful discussions.

%\bibliographystyle{apsrev}
%\bibliography{prisoner}

\begin{thebibliography}{26}
\expandafter\ifx\csname natexlab\endcsname\relax\def\natexlab#1{#1}\fi
\expandafter\ifx\csname bibnamefont\endcsname\relax
  \def\bibnamefont#1{#1}\fi
\expandafter\ifx\csname bibfnamefont\endcsname\relax
  \def\bibfnamefont#1{#1}\fi
\expandafter\ifx\csname citenamefont\endcsname\relax
  \def\citenamefont#1{#1}\fi
\expandafter\ifx\csname url\endcsname\relax
  \def\url#1{\texttt{#1}}\fi
\expandafter\ifx\csname urlprefix\endcsname\relax\def\urlprefix{URL }\fi
\providecommand{\bibinfo}[2]{#2}
\providecommand{\eprint}[2][]{\url{#2}}

\bibitem[{\citenamefont{Darwin}(1859)}]{darwin}
\bibinfo{author}{\bibfnamefont{C.~R.} \bibnamefont{Darwin}},
  \emph{\bibinfo{title}{On The Origin of Species by Means of Natural
  Selection}} (\bibinfo{publisher}{John Murray}, \bibinfo{address}{London},
  \bibinfo{year}{1859}).

\bibitem[{\citenamefont{Dawkins}(1976)}]{dawkins}
\bibinfo{author}{\bibfnamefont{R.}~\bibnamefont{Dawkins}},
  \emph{\bibinfo{title}{The Selfish Gene}} (\bibinfo{publisher}{Oxford
  University Press}, \bibinfo{year}{1976}).

\bibitem[{\citenamefont{Boorman and Levitt}(1980)}]{boorman}
\bibinfo{author}{\bibfnamefont{S.~A.} \bibnamefont{Boorman}} \bibnamefont{and}
  \bibinfo{author}{\bibfnamefont{P.~R.} \bibnamefont{Levitt}},
  \emph{\bibinfo{title}{The genetics of altruism}}
  (\bibinfo{publisher}{Academic}, \bibinfo{address}{New York},
  \bibinfo{year}{1980}).

\bibitem[{\citenamefont{Hardin}(1968)}]{hardin_1968}
\bibinfo{author}{\bibfnamefont{G.}~\bibnamefont{Hardin}},
  \bibinfo{journal}{Science} \textbf{\bibinfo{volume}{162}},
  \bibinfo{pages}{1243} (\bibinfo{year}{1968}).

\bibitem[{\citenamefont{Nowak and May}(1992)}]{nowak1992}
\bibinfo{author}{\bibfnamefont{M.~A.} \bibnamefont{Nowak}} \bibnamefont{and}
  \bibinfo{author}{\bibfnamefont{R.~M.} \bibnamefont{May}},
  \bibinfo{journal}{Nature} \textbf{\bibinfo{volume}{359}},
  \bibinfo{pages}{826} (\bibinfo{year}{1992}).

\bibitem[{\citenamefont{Axelrod}(1984)}]{axelrod1984}
\bibinfo{author}{\bibfnamefont{R.}~\bibnamefont{Axelrod}},
  \emph{\bibinfo{title}{The evolution of cooperation}}
  (\bibinfo{publisher}{Basic Books}, \bibinfo{address}{New York},
  \bibinfo{year}{1984}).

\bibitem[{\citenamefont{Schweitzer et~al.}(2002)\citenamefont{Schweitzer,
  Behera, and Muhlenbein}}]{schweitzer_2002}
\bibinfo{author}{\bibfnamefont{F.}~\bibnamefont{Schweitzer}},
  \bibinfo{author}{\bibfnamefont{L.}~\bibnamefont{Behera}}, \bibnamefont{and}
  \bibinfo{author}{\bibfnamefont{H.}~\bibnamefont{Muhlenbein}},
  \bibinfo{journal}{Advances in Complex Systems} \textbf{\bibinfo{volume}{5}},
  \bibinfo{pages}{269} (\bibinfo{year}{2002}),
  \bibinfo{note}{cond-mat/0211605}.

\bibitem[{\citenamefont{Axelrod and Hamilton}(1981)}]{axelrod1981}
\bibinfo{author}{\bibfnamefont{R.}~\bibnamefont{Axelrod}} \bibnamefont{and}
  \bibinfo{author}{\bibfnamefont{W.~D.} \bibnamefont{Hamilton}},
  \bibinfo{journal}{Science} \textbf{\bibinfo{volume}{211}},
  \bibinfo{pages}{1390} (\bibinfo{year}{1981}).

\bibitem[{\citenamefont{Nowak and Sigmund}(1993)}]{nowak1993}
\bibinfo{author}{\bibfnamefont{M.~A.} \bibnamefont{Nowak}} \bibnamefont{and}
  \bibinfo{author}{\bibfnamefont{K.}~\bibnamefont{Sigmund}},
  \bibinfo{journal}{Nature} \textbf{\bibinfo{volume}{364}}, \bibinfo{pages}{56}
  (\bibinfo{year}{1993}).

\bibitem[{\citenamefont{Nowak et~al.}(1994)\citenamefont{Nowak, Bonhoeffer, and
  May}}]{nowak1994}
\bibinfo{author}{\bibfnamefont{M.~A.} \bibnamefont{Nowak}},
  \bibinfo{author}{\bibfnamefont{S.}~\bibnamefont{Bonhoeffer}},
  \bibnamefont{and} \bibinfo{author}{\bibfnamefont{R.~M.} \bibnamefont{May}},
  \bibinfo{journal}{Int. J. Chaos Bifurc.} \textbf{\bibinfo{volume}{4}},
  \bibinfo{pages}{33} (\bibinfo{year}{1994}).

\bibitem[{\citenamefont{Silva and Fontanari}(1999)}]{fontanari_epjb_1999}
\bibinfo{author}{\bibfnamefont{A.~T.~C.} \bibnamefont{Silva}} \bibnamefont{and}
  \bibinfo{author}{\bibfnamefont{J.~F.} \bibnamefont{Fontanari}},
  \bibinfo{journal}{Eur. Phys. J. B} \textbf{\bibinfo{volume}{7}},
  \bibinfo{pages}{385} (\bibinfo{year}{1999}).

\bibitem[{\citenamefont{Alves et~al.}(2000)\citenamefont{Alves, Campos, Silva,
  and Fontanari}}]{fontanari_pre_2000}
\bibinfo{author}{\bibfnamefont{D.}~\bibnamefont{Alves}},
  \bibinfo{author}{\bibfnamefont{P.~R.~A.} \bibnamefont{Campos}},
  \bibinfo{author}{\bibfnamefont{A.~T.~C.} \bibnamefont{Silva}},
  \bibnamefont{and} \bibinfo{author}{\bibfnamefont{J.~F.}
  \bibnamefont{Fontanari}}, \bibinfo{journal}{Phys. Rev. E}
  \textbf{\bibinfo{volume}{63}}, \bibinfo{pages}{011911}
  (\bibinfo{year}{2000}).

\bibitem[{\citenamefont{Kim et~al.}(2002)\citenamefont{Kim, Trusina, Holme,
  Minnhagen, Chung, and Choi}}]{kim}
\bibinfo{author}{\bibfnamefont{B.~J.} \bibnamefont{Kim}},
  \bibinfo{author}{\bibfnamefont{A.}~\bibnamefont{Trusina}},
  \bibinfo{author}{\bibfnamefont{P.}~\bibnamefont{Holme}},
  \bibinfo{author}{\bibfnamefont{P.}~\bibnamefont{Minnhagen}},
  \bibinfo{author}{\bibfnamefont{J.~S.} \bibnamefont{Chung}}, \bibnamefont{and}
  \bibinfo{author}{\bibfnamefont{M.~Y.} \bibnamefont{Choi}},
  \bibinfo{journal}{Phys. Rev. E} \textbf{\bibinfo{volume}{66}},
  \bibinfo{pages}{1} (\bibinfo{year}{2002}).

\bibitem[{\citenamefont{Vainstein and Arenzon}(2001)}]{vainstein_2001}
\bibinfo{author}{\bibfnamefont{M.~H.} \bibnamefont{Vainstein}}
  \bibnamefont{and} \bibinfo{author}{\bibfnamefont{J.~J.}
  \bibnamefont{Arenzon}}, \bibinfo{journal}{Phys. Rev. E}
  \textbf{\bibinfo{volume}{64}}, \bibinfo{pages}{051905}
  (\bibinfo{year}{2001}).

\bibitem[{\citenamefont{D\'uran and Mulet}()}]{duran_2003}
\bibinfo{author}{\bibfnamefont{O.}~\bibnamefont{D\'uran}} \bibnamefont{and}
  \bibinfo{author}{\bibfnamefont{R.}~\bibnamefont{Mulet}},
  \bibinfo{note}{cond-mat/0305353}.

\bibitem[{\citenamefont{Greig and Trevisano}(2004)}]{greig}
\bibinfo{author}{\bibfnamefont{D.}~\bibnamefont{Greig}} \bibnamefont{and}
  \bibinfo{author}{\bibfnamefont{M.}~\bibnamefont{Trevisano}},
  \bibinfo{journal}{Proc. R. Soc. Lond. B (Suppl.)}
  \textbf{\bibinfo{volume}{271}}, \bibinfo{pages}{S25} (\bibinfo{year}{2004}).

\bibitem[{\citenamefont{Turner and Chao}(1998)}]{turner1}
\bibinfo{author}{\bibfnamefont{P.~E.} \bibnamefont{Turner}} \bibnamefont{and}
  \bibinfo{author}{\bibfnamefont{L.}~\bibnamefont{Chao}},
  \bibinfo{journal}{Genetics} \textbf{\bibinfo{volume}{150}},
  \bibinfo{pages}{523} (\bibinfo{year}{1998}).

\bibitem[{\citenamefont{Turner and Chao}(1999)}]{turner2}
\bibinfo{author}{\bibfnamefont{P.~E.} \bibnamefont{Turner}} \bibnamefont{and}
  \bibinfo{author}{\bibfnamefont{L.}~\bibnamefont{Chao}},
  \bibinfo{journal}{Nature} \textbf{\bibinfo{volume}{398}},
  \bibinfo{pages}{441} (\bibinfo{year}{1999}).

\bibitem[{\citenamefont{Dugatkin}(1989)}]{dugatkin_1989}
\bibinfo{author}{\bibfnamefont{L.~A.} \bibnamefont{Dugatkin}},
  \bibinfo{journal}{J. Theoretical Biology} \textbf{\bibinfo{volume}{142}},
  \bibinfo{pages}{123} (\bibinfo{year}{1989}).

\bibitem[{\citenamefont{Dugatkin}(1991)}]{dugatkin_1991}
\bibinfo{author}{\bibfnamefont{L.~A.} \bibnamefont{Dugatkin}},
  \bibinfo{journal}{Behav. Ecol. Sociobiol.} \textbf{\bibinfo{volume}{29}},
  \bibinfo{pages}{127} (\bibinfo{year}{1991}).

\bibitem[{\citenamefont{Milinski}(1990)}]{milinski}
\bibinfo{author}{\bibfnamefont{M.}~\bibnamefont{Milinski}},
  \bibinfo{journal}{Anim. Behav.} \textbf{\bibinfo{volume}{40}},
  \bibinfo{pages}{1190} (\bibinfo{year}{1990}).

\bibitem[{\citenamefont{Smith}(1982)}]{smith_1982}
\bibinfo{author}{\bibfnamefont{J.~M.} \bibnamefont{Smith}},
  \emph{\bibinfo{title}{Evolution and the theory of games}}
  (\bibinfo{publisher}{Cambridge University Press},
  \bibinfo{address}{Cambridge}, \bibinfo{year}{1982}).

\bibitem[{\citenamefont{Stauffer}(2004)}]{staufer_2004}
\bibinfo{author}{\bibfnamefont{D.}~\bibnamefont{Stauffer}},
  \bibinfo{journal}{Physica A} \textbf{\bibinfo{volume}{336}},
  \bibinfo{pages}{1} (\bibinfo{year}{2004}).

\bibitem[{\citenamefont{Galam}(2004)}]{galam_2004}
\bibinfo{author}{\bibfnamefont{S.}~\bibnamefont{Galam}},
  \bibinfo{journal}{Physica A} \textbf{\bibinfo{volume}{336}},
  \bibinfo{pages}{49} (\bibinfo{year}{2004}).

\bibitem[{\citenamefont{Bouchaud}(2002)}]{bouchaud_2002}
\bibinfo{author}{\bibfnamefont{J.-P.} \bibnamefont{Bouchaud}},
  \bibinfo{journal}{Physica A} \textbf{\bibinfo{volume}{313}},
  \bibinfo{pages}{238} (\bibinfo{year}{2002}).

\bibitem[{\citenamefont{Anteneodo et~al.}(2002)\citenamefont{Anteneodo,
  Tsallis, and Martinez}}]{anteneodo:1:2002}
\bibinfo{author}{\bibfnamefont{C.}~\bibnamefont{Anteneodo}},
  \bibinfo{author}{\bibfnamefont{C.}~\bibnamefont{Tsallis}}, \bibnamefont{and}
  \bibinfo{author}{\bibfnamefont{A.~S.} \bibnamefont{Martinez}},
  \bibinfo{journal}{Europhys. Lett.} \textbf{\bibinfo{volume}{59}},
  \bibinfo{pages}{635} (\bibinfo{year}{2002}).

\end{thebibliography}

\end{document}